\documentclass[aps,prl,twocolumn,superscriptaddress,floatfix,longbibliography]{revtex4-2}
\usepackage[utf8]{inputenc}
\usepackage[T1]{fontenc}
\usepackage{amsmath, amssymb, graphicx, bm}
\usepackage{braket}
\usepackage[dvipsnames]{xcolor}
\usepackage[colorlinks=true, citecolor=blue, urlcolor=blue, linkcolor=blue, breaklinks=true, pdfpagelabels=false, hypertexnames=false]{hyperref}
\usepackage{booktabs}
\usepackage{array}

\makeatletter
\def\selectlanguage#1{}
\makeatother

\newcommand{\rs}{r_s}
\newcommand{\eh}{electron--hole}
\newcommand{\la}{\langle}
\newcommand{\ra}{\rangle}


\begin{document}

\title{Honeycomb Crystal in a Mass-Asymmetric Electron-Hole Bilayer}

\author{Shengdu Chai}
\affiliation{Interdisciplinary Center for Theoretical Physics and Information Sciences (ICTPIS), Fudan University, Shanghai 200433, China}
\affiliation{Shanghai Artificial Intelligence Laboratory, Shanghai 200232, China}
\author{Wangqian Miao}
\affiliation{Department of Physics, The Hong Kong University of Science and Technology, Clear Water Bay, Hong Kong, China}
\author{Yuqiang Li}
\affiliation{Shanghai Artificial Intelligence Laboratory, Shanghai 200232, China}
\author{Lei Wang}
\affiliation{Beijing National Laboratory for Condensed Matter Physics and Institute of Physics, Chinese Academy of Sciences, Beijing 100190, China}
\author{Xi Dai}
\affiliation{Department of Physics, The Hong Kong University of Science and Technology, Clear Water Bay, Hong Kong, China}
\author{X.C. Xie}
\affiliation{Interdisciplinary Center for Theoretical Physics and Information Sciences (ICTPIS), Fudan University, Shanghai 200433, China}
\affiliation{International Center for Quantum Materials, School of Physics, Peking University, Beijing 100871, China}
\affiliation{Hefei National Laboratory, Hefei 230088, China}
\date{\today}

\begin{abstract}
    A two-dimensional electron gas in a uniform neutralizing background forms a triangular Wigner crystal at low density.
    In a strongly mass-asymmetric \eh{} bilayer, localized heavy holes instead provide a nonuniform positive background that supports an electron crystal at densities two orders of magnitude higher than for a homogeneous electron gas in the same material.
    Here we show that correlated mobile electrons can overturn the preference for triangular order, stabilizing a honeycomb \eh{} crystal.
    Using the powerful real-space neural-network variational Monte Carlo in the Born--Oppenheimer limit with fixed holes, we find that increasing the density drives a transition from a triangular \eh{} crystal to a honeycomb crystal, which eventually melts into itinerant electrons coexisting with the honeycomb or the triangular hole lattice at fixed interlayer separation.
    Relative to triangular order, honeycomb order emerges when gains in electronic kinetic energy and \eh{} attraction outweigh its higher intralayer Coulomb repulsion, a threshold never reached in the rigid point-dipole model.
    Moreover, the honeycomb hole lattice supports paramagnetic, antiferromagnetic, and stripe-ordered spin states.
    Complex polarization indicates a metal--insulator transition, while the states on the honeycomb hole lattice are consistent with a stripe-ordered Dirac semimetal and paramagnetic or antiferromagnetic Mott insulators via mapping to the half-filled Hubbard model.
    We further propose a BLG/hBN/WS$_2$--WSe$_2$ heterostructure as a platform for the predicted honeycomb crystal, at an experimentally feasible density $n\sim10^{11}~\text{cm}^{-2}$.
\end{abstract}

\maketitle

\setcounter{secnumdepth}{0}

\paragraph{Introduction}
As one of the earliest theoretical predictions in many-electron physics, the Wigner crystal forms when long-range Coulomb repulsion overwhelms the kinetic energy of a dilute electron gas~\cite{wigner_interaction_1934}, a competition later quantified by quantum Monte Carlo~\cite{zhu_wigner_1995,drummond_low_density_2009,azadi_phase_2024}.
Experiments first found evidence for a two-dimensional electron crystal in strong magnetic fields~\cite{andrei_observation_1988,goldman_evidence_1990}.
More recently, experiments have reported evidence for electronic Wigner crystals in transition-metal dichalcogenide (TMD) layers~\cite{zhou_bilayer_2021}, cadmium arsenide thin films~\cite{munyan_evidence_2024}, and rhombohedral graphene multilayers~\cite{han_evidence_2026,zhou_competing_2026,sun2026realspacevisualizationemergentelectron}.

For a conventional two-dimensional Wigner crystal, classical electrostatics favors triangular order~\cite{bonsall_static_1977}
, a preference for which a general proof has recently been proposed~\cite{OAI:An-atomic-certificate-for-triangular-lattice-universal-optimality-September-26-2026,OAI:Universal-optimality-of-the-triangular-lattice-September-23-2026}
.
This classical preference, however, relies on treating the compensating charge as a uniform background.
A neutral \eh{} bilayer is qualitatively different: with both charge components explicit, the electronic ground-state energy depends directly on the hole-lattice geometry.
We focus on a strongly mass-asymmetric regime in which the effective hole mass greatly exceeds the electron mass, yielding strongly localized holes and relatively itinerant electrons.

Spatially separated \eh{} bilayers provide an experimental setting for this hierarchy.
Such bilayers host strongly correlated excitonic physics identified by different experimental probes in TMD systems~\cite{ma_strongly_2021,qi_thermodynamic_2023, qi_perfect_2025,nguyen_perfect_2025}.
A recent device combined electrons in a nonmoiré MoSe$_2$ monolayer with holes in a WS$_2$/WSe$_2$ moiré superlattice~\cite{qi_exciton_2026}.
Because the moiré potential acts only on the hole layer, it suppresses hole motion relative to the nonmoiré electrons, while strong correlations further localize the holes at commensurate fillings, where they can form triangular, honeycomb, and kagome lattices~\cite{pan_quantum_2020, li_imaging_2021}.
These developments motivate our investigation of a mass-asymmetric Coulomb bilayer, where mobile correlated electrons couple to a static hole crystal.

Theoretical studies of continuum \eh{} bilayers have identified excitonic condensates, dipolar superfluids, and Wigner crystals across density, interlayer separation, and mass imbalance~\cite{depalo_excitonic_2002,balatsky_dipolar_2004,joglekar_wigner_2006,schleede_phase_2012,maezono_excitons_2013,neilson_excitonic_2014,sharma_ground_2016,moroni_quadriexcitons_2023,dai_strong_2024}.
However, whether mobile electrons can reconstruct the hole lattice remains an open question.
Answering this question requires a flexible representation of the mobile electrons alongside accurate calculations of small energy differences between competing hole lattices.
Real-space neural-network variational Monte Carlo (NN-VMC) is well suited to this task; it captures competing electronic states and yields both high-accuracy variational energies and real-space charge and spin densities.
NN-VMC has been applied broadly to strongly correlated lattice and continuum models~\cite{carleo_solving_2017,pfau_abinitio_2020,hermann_initio_2023,qian_deep_2025,han_solving_2019,Hermann:2020xqs,Pfau:2023azx,foster_initio_2025,li_spinadapted_2025,li_ab_initio_2022,li_emergent_2024,luo_simulating_2024,smith_neural_2026,lou_neural_2024,li_attention_2025,xie_deep_2023a,dong_deep_2025,linteau_universal_2025,chai_revisiting_2025,qian_describing_2025,teng_solving_2025,gattu_dressing_2025a,abouelkomsan_topological_2025a,li_deep_2025}.
In particular, it has described liquid, crystalline, excitonic, and paired states in electron gases and \eh{} bilayers~\cite{wilson_neural_2023,cassella_discovering_2023,pescia_message_2024,smith_ground_2024,valenti_quantum_2025,ge_visualizing_2025,luo_pairing-based_2026}.

In this Letter, we demonstrate that mobile electrons can indeed reconstruct the hole lattice.
By independently optimizing the electronic ground state for triangular (T), honeycomb (HC), and kagome (KG) hole lattices at fixed pair density, we show that the electronic energy can overcome the classical advantage of triangular order, stabilizing a honeycomb lattice at intermediate densities.
Moreover, mass asymmetry extends electronic charge order to densities nearly two orders of magnitude higher than for Wigner crystallization in a homogeneous electron gas.
A total-energy decomposition shows that the honeycomb phase is stabilized by a combined gain in electronic kinetic and interlayer Coulomb energies that outweighs the electron--electron and hole--hole penalties.
This gain can be attributed to the shorter honeycomb bonds, which enhance hopping and double occupancy, suppressing double occupancy restores triangular order.
This reconstruction is therefore a quantum many-body effect: it vanishes in a rigid point-dipole reference where each electron is replaced by a point charge fixed directly above a hole~\cite{OAI:Universal-optimality-of-the-triangular-lattice-September-23-2026,OAI:An-atomic-certificate-for-triangular-lattice-universal-optimality-September-26-2026,skinner_anomalously_2010a}.
For the honeycomb hole lattice, mapping the optimized states onto an effective half-filled Hubbard model exposes a range of electronic behavior, including a stripe-ordered Dirac semimetal and paramagnetic (PM) and antiferromagnetic (AFM) Mott insulators.
We characterize the associated metal--insulator transition using the complex polarization~\cite{resta_quantum_1998,resta_electron_1999,motta_ground_2020}.
These results show how quantum many-body effects determine both the structural geometry and the electronic character of mass-asymmetric \eh{} bilayers.
\begin{figure}[t]
    \centering
    \includegraphics[width=\columnwidth]{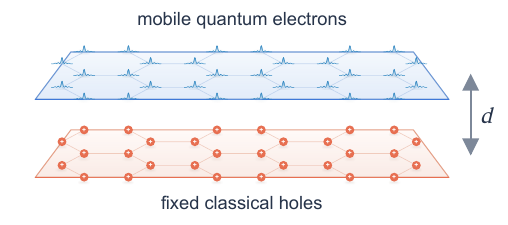}
    \caption{Fixed-hole model for a charge-neutral, mass-asymmetric \eh{} bilayer with interlayer separation $d$: $N$ mobile quantum electrons (blue) and $N$ classical holes fixed as point charges in the Born--Oppenheimer limit (orange).}
    \label{fig:model_setup}
\end{figure}
\paragraph{Model and method}
We model the system as a charge-neutral, two-dimensional \eh{} bilayer with an interlayer separation $d$ [Fig.~\ref{fig:model_setup}].
Taking the Born--Oppenheimer (BO) limit $m_h/m_e\to\infty$ freezes the $N$ holes at positions $\{\mathbf h_I\}$, while the $N$ electrons at positions $\{\mathbf e_i\}$ remain the mobile quantum degrees of freedom.
This BO treatment is conceptually related to one-, two-, and three-dimensional hydrogen systems~\cite{motta_ground_2020,calcavecchia_twodimensional_2017,natoliCrystalStructureAtomic1993,cozza_lowsymmetry_2026}, with the fixed holes playing the role of protons.
We adopt Hartree atomic units, where the Bohr radius $a_{\mathrm{B}}=4\pi\epsilon_0\hbar^{2}/(m_e e^{2})$ is the unit of length and the Hartree energy $E_{\mathrm{Ha}}=e^{2}/(4\pi\epsilon_0a_{\mathrm{B}})$ is the unit of energy.
In these units, the fixed-hole Hamiltonian is
\begin{align}
    \hat{H}={} & -\frac{1}{2}\sum_i\nabla_{\mathbf e_i}^2-\sum_{i,I}\frac{1}{\sqrt{|\mathbf e_i-\mathbf h_I|^2+d^2}}
    \nonumber                                                                                                        \\& +\frac{1}{2}\sum_{i\ne j}\frac{1}{|\mathbf e_i-\mathbf e_j|}
    +\frac{1}{2}\sum_{I\ne J}\frac{1}{|\mathbf h_I-\mathbf h_J|}.
    \label{eq:hamiltonian}
\end{align}
We parameterize the average electron density by the dimensionless Wigner--Seitz radius $\rs=1/\sqrt{\pi n}$, where $n=N/A$ and the area $A$ is measured in $a_{\mathrm{B}}^2$.
We consider the spin-unpolarized sector with $N_{\uparrow}=N_{\downarrow}=N/2$, while the holes are treated as spinless classical charges.

At fixed density and interlayer separation, the ground-state hole lattice is the geometry that minimizes the BO energy among the candidate lattices.
For each candidate hole lattice, the total energy separates exactly
into the classical hole--hole and quantum electronic contributions,
$E_{\mathrm{BO}}=V_{hh}+E_{\mathrm{el}}$, where
$E_{\mathrm{el}}=\min_{\Psi}
    \braket{\Psi|\hat T+\hat V_{ee}+\hat V_{eh}|\Psi}$.
All long-range periodic Coulomb interactions are evaluated using the two-dimensional Parry Ewald summation~\cite{parry_electrostatic_1975,parry_errata_1976,mazars_lekner_2005,lindbo_fast_2012} (Supplemental Material (SM)~\cite{supplemental} Sec.~S~III.C).

We estimate the ground-state electronic energy of each candidate lattice using a real-space NN-VMC ansatz of the Slater-determinant form
\begin{equation}
    \Psi(\mathbf e;\mathbf h)
    =
    \sum_{\nu}
    \det[\phi_{\nu}],
    \label{eq:nn_vmc_ansatz}
\end{equation}
where $[\phi_{\nu}]_{ip}=\phi_{\nu p}(\mathbf e_i;\mathbf e,\mathbf h)$.
Each determinant enforces fermionic antisymmetry by construction.
We construct the generalized orbitals $\phi_{\nu p}$ by integrating self-attention mechanisms~\cite{pfau_self-attention_2023} with periodic electron-gas descriptors~\cite{wilson_neural_2023}.
To adapt this framework to the \eh{} bilayer, we introduce a softened, hole-centered envelope function that suppresses short-distance kinetic singularities (details are provided in SM~\cite{supplemental}).
\begin{figure}[t]
    \centering
    \includegraphics[width=\columnwidth]{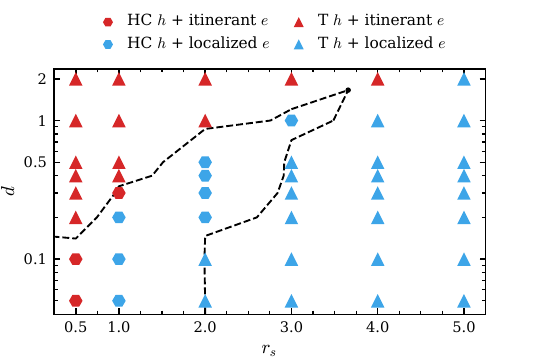}
    \caption{Phase diagram of the ground-state hole lattice and electronic character for the $N=36$ spin-unpolarized ($N_{\uparrow}=N_{\downarrow}=18$) system. Marker shapes give the ground-state hole lattice (hexagons, HC; triangles, T), and colors give the electronic character (red, itinerant; blue, localized). Dashed lines show the interpolated boundaries from the energy comparison between hole lattices. Finite-size checks at $N=48$ near the phase boundaries support the phase diagram, with the inferred boundaries remaining nearly unchanged.
    }
    \label{fig:structural_sequence}
\end{figure}
\paragraph{Charge and spin order}

\begin{figure*}[ht]
    \centering
    \includegraphics[width=\textwidth]{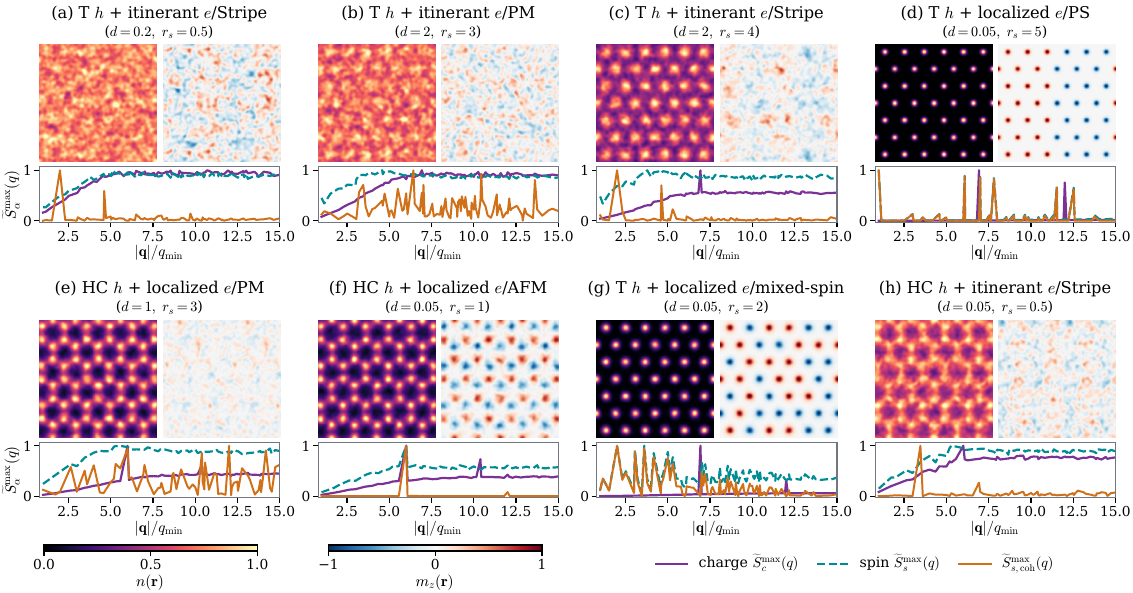}
    \caption{
        Charge and spin phases of the optimized $N=36$ states. The upper panels show the real-space charge $n(\mathbf r)$ (left) and spin $m_z(\mathbf r)$ (right) for each state, with each panel normalized independently. The lower panels plot the normalized radial-shell maxima $\widetilde S_c^{\max}(q)$ (purple solid), $\widetilde S_s^{\max}(q)$ (teal dashed) and $\widetilde S_{s,\rm coh}^{\max}(q)$ (orange solid) against $q/q_{\min}$ [Eqs.~(\ref{eq:charge_spin_structure})--(\ref{eq:radial_shell_spectra})]. Each panel title gives the hole lattice [triangular (T) or honeycomb (HC)], the electronic character (itinerant or localized), and the spin state.}
    \label{fig:magnetic_textures}
\end{figure*}

\begin{figure*}[ht]
    \centering
    \includegraphics[width=\textwidth]{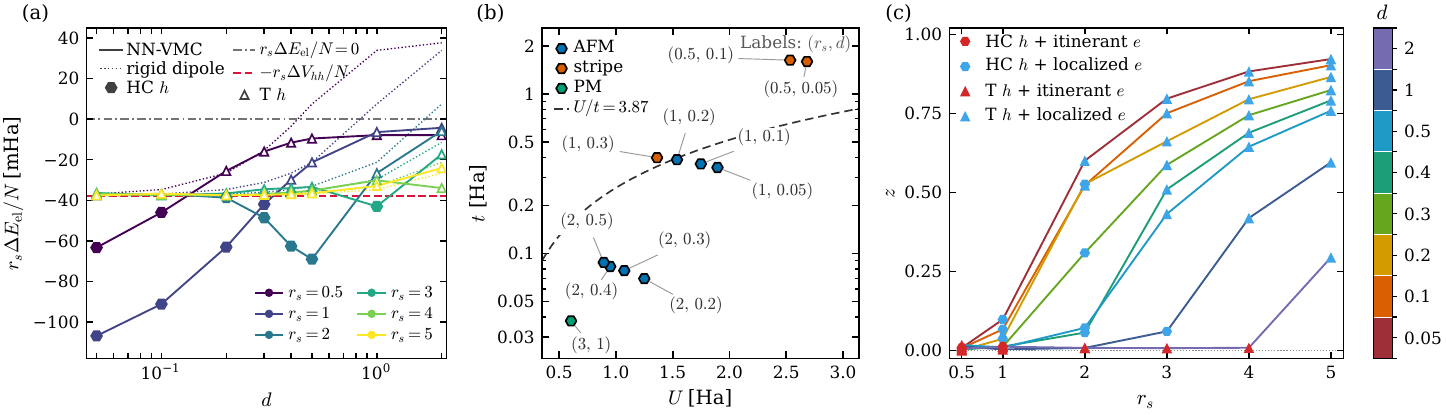}
    \caption{
        Mechanism and electronic properties of the optimized $N=36$ states.
        (a) Scaled electronic energy difference $\rs\Delta E_{\mathrm{el}}/N$ between honeycomb and triangular states.
        Solid and dotted curves show the NN-VMC and rigid point-dipole results, respectively.
        The red dashed line marks the gain required to overcome the hole--hole penalty, $-\rs\Delta V_{hh}/N=-37.7$~mHa.
        Marker shapes give the ground-state hole lattice, and line colors give $\rs$.
        (b) Nearest-neighbor hopping $t$ versus on-site repulsion $U$ for the ground states on the honeycomb hole lattice, with a logarithmic vertical axis.
        Colors distinguish the AFM, stripe, and PM states, and point labels give $(\rs,d)$.
        The dashed line is $U/t=3.87$~\cite{sorella_absence_2012,otsuka_universal_2016}.
        (c) Complex polarization $z$ of the electrons ~\cite{resta_quantum_1998,resta_electron_1999}.
        Marker shapes give the hole lattice and marker colors the electronic character (red, itinerant; blue, localized); line colors give $d$ (color bar).
    }
    \label{fig:stability_electronic_properties}
\end{figure*}

Fig.~\ref{fig:structural_sequence} labels each calculated point of the $N=36$ spin-unpolarized system by the ground-state hole lattice and by whether the electrons are localized or itinerant.
We denote localized electrons on a triangular (honeycomb) hole lattice as a triangular (honeycomb) \eh{} crystal.
We distinguish localized from itinerant electrons by the complex polarization~\cite{resta_quantum_1998,resta_electron_1999,motta_ground_2020} for the electrons (as shown in Fig.~\ref{fig:stability_electronic_properties}(c)).
At fixed $d$, increasing the density (decreasing $\rs$) drives the system from the triangular crystal to the honeycomb crystal, and then to itinerant electrons.
While the triangular geometry dominates at low density, the electronic contribution lowers the relative energy of the honeycomb lattice, stabilizing it over an intermediate-density regime.
Kagome is never the ground-state hole lattice in the parameter range studied.
Calculations at $N=48$ near the phase boundaries reproduce the $N=36$ phase assignments, with only marginal shifts in the inferred boundaries. The energy comparisons are shown in SM~\cite{supplemental}.

As $d$ increases, the boundary between the charge-ordered and itinerant electron states shifts to larger $\rs$, from $\rs<1$ at small $d$ to $4<\rs<5$ at large $d$.
This shift follows from the weakening periodic potential produced by the hole lattice, whose Fourier component at reciprocal-lattice wave vector $\mathbf G\ne0$ decays as $e^{-|\mathbf G|d}$.
In the $d\to\infty$ limit, this potential vanishes and the electron layer recovers a homogeneous two-dimensional electron gas, for which triangular Wigner crystallization occurs at $\rs\simeq31$--$37$~\cite{drummond_low_density_2009,azadi_phase_2024,smith_ground_2024}.
At finite $d$, the hole lattice instead localizes electron charge order already at $\rs\lesssim4$, roughly one order of magnitude below the homogeneous-electron-gas threshold.
Since $n\propto\rs^{-2}$, this reduction raises the corresponding density by nearly two orders of magnitude, making the ordered regime substantially more accessible experimentally.

While the charge-order diagram identifies whether the electrons crystallize and which hole lattice they select, it leaves the spin degrees of freedom unspecified.
Extracting the spin densities of the optimized states gives five distinct spin states: stripe, PM, AFM, mixed-spin, and phase-separated (PS) [Fig.~\ref{fig:magnetic_textures}].
Depending on $d$, increasing the density drives the system through one of two distinct phase progressions.
In one sequence, the system transitions from the PS triangular crystal to the mixed-spin triangular crystal, followed by the AFM honeycomb crystal, and ultimately reaches a stripe state on either the honeycomb or the triangular hole lattice.
In the other sequence, the system proceeds from the PS triangular crystal into a stripe state on the triangular hole lattice or the PM honeycomb crystal, then into a PM state with itinerant electrons on the triangular hole lattice, and eventually terminates in a stripe state on the triangular hole lattice.

For each of these charge-spin states, the $n(\mathbf r)$ and $m_z(\mathbf r)$ panels display the charge- and spin-density patterns, while the radial structure factors identify their ordering wave vectors and periodicities.
Notably, the triangular crystal with PS shows a ferromagnetic tendency within the constrained $S_z=0$ sector, where spin-up and spin-down electrons segregate into macroscopic domains [Fig.~\ref{fig:magnetic_textures}(d)].
The full charge- and spin-density patterns are provided in SM~\cite{supplemental}.

\paragraph{Mechanism}

To isolate the mechanism driving the hole-lattice reconstruction, we decompose the total energy difference between the honeycomb and triangular phases into the classical hole--hole repulsion and the quantum electronic energy, defining $\Delta X/N=(X_{\mathrm{HC}}-X_{\mathrm{T}})/N$.
Because the holes are fixed, their classical repulsion $V_{hh}$ in Eq.~(\ref{eq:hamiltonian}) simplifies to a purely structural two-dimensional Madelung summation that is independent of the interlayer separation $d$. Writing the Madelung energy per hole as $-\alpha/\rs$, we obtain
$
    \frac{\rs\,\Delta V_{hh}}{N}=\alpha_{\mathrm{T}}-\alpha_{\mathrm{HC}}
    \simeq 37.7~\text{mHa}.
$
The positive difference represents the classical energy penalty for reconstructing the hole lattice from triangular to honeycomb order. In contrast, the NN-VMC electronic contribution favors the honeycomb lattice throughout the investigated parameter range ($\Delta E_{\mathrm{el}}<0$), as shown in Fig.~\ref{fig:stability_electronic_properties}(a).
This electronic energy gain, however, is not always sufficient to compensate for the classical penalty.
The total energy stabilizes honeycomb order when $\rs\Delta E_{\mathrm{el}}/N <-\rs\Delta V_{hh}/N\simeq-37.7~\mathrm{mHa}$.
For the classical rigid point-dipole reference, the electronic-energy difference stays above this threshold over the same range.

A generalized Hubbard model within the Gutzwiller approximation~\cite{gutzwiller_effect_1963,vollhardt_normal_1984} provides insight into the emergence of the honeycomb hole lattice at intermediate densities (details are provided in SM~\cite{supplemental}).
At the same density, the shorter bonds in the honeycomb geometry enhance orbital overlap and hence the hopping scale.
Consequently, the honeycomb lattice exhibits a higher double occupancy $D$ and achieves a stronger one-body energy gain (including kinetic energy and \eh{} attraction) than its triangular counterpart.
This mechanism is further supported by the NN-VMC energy decomposition shown in SM Fig.~S8~\cite{supplemental}, where the combined kinetic and interlayer Coulomb contributions favor the honeycomb structure, whereas the intralayer Coulomb terms favor the triangular order.
The honeycomb lattice becomes the ground-state lattice when this additional gain exceeds the Coulomb repulsion.
A finite double occupancy is essential: constraining $D=0$ removes this one-body energy gain, and the triangular lattice is then always lower in energy.
Also, when the hole lattice is replaced by a uniform positive background, as in the two-dimensional electron gas, the triangular lattice remains lower in energy even with $D$ optimized.

Moreover, the honeycomb geometry may lower its energy further through a Kekulé distortion or dimerization~\cite{naumov_graphene_2013,motta_ground_2020}, in analogy with hydrogen systems~\cite{naumov_graphene_2013,ceperleyGroundStateSolid1987,Monacelli:2022cex,motta_ground_2020,cozza_lowsymmetry_2026,calcavecchia_twodimensional_2017,natoliCrystalStructureAtomic1993}.

\paragraph{Electronic properties} Since our honeycomb hole lattice hosts one electron per site, it can be mapped onto the half-filled honeycomb Hubbard model, which undergoes a Dirac-semimetal to Mott-insulator transition at $U/t \simeq 3.87$~\cite{sorella_absence_2012,otsuka_universal_2016} (see Appendix~\ref{app:electronic_character} for details). As shown in Fig.~\ref{fig:stability_electronic_properties}(b), the stripe states on the honeycomb hole lattice lie below this threshold and are consistent with a Dirac-semimetal description, whereas the AFM and PM honeycomb crystals lie above it, as expected for correlated Mott insulators.

We use the optimized wave functions to compute the magnitude of the complex polarization $z=|\la e^{i\mathbf b\cdot\sum_j\mathbf e_j}\ra|$ as a complementary signature of electronic localization~\cite{resta_quantum_1998,resta_electron_1999,motta_ground_2020}, with $\mathbf b$ a chosen reciprocal vector of the simulation cell.
The metal--insulator transition is governed by both the interaction strength $\rs$ and the interlayer separation $d$: increasing $\rs$ stabilizes the insulating state (finite $z$), whereas a larger $d$ shifts its onset to higher $\rs$.
As shown in Fig.~\ref{fig:stability_electronic_properties}(c), the stripe states on the honeycomb hole lattice exhibit near-zero $z$, supporting their assignment as semimetal phases in the Hubbard model.
The AFM and PM honeycomb crystals display a finite value of $z$, which also supports the Mott-insulator description in the Hubbard model.
On the triangular hole lattice, the stripe and PM states yield near-zero $z$, as expected for a metal, whereas the PS and mixed-spin triangular crystals show finite $z$, consistent with insulating behavior.

\paragraph{Discussion and conclusions}
Our NN-VMC calculations demonstrate that mobile electrons can drive a reconstruction of the ground-state hole lattice in strongly mass-asymmetric Coulomb bilayers. We find that the interplay of electronic kinetic and interlayer Coulomb energy gains can overcome the intralayer Coulomb penalties, robustly stabilizing a honeycomb crystal across an intermediate-density regime.
Previous work~\cite{nashabeh_prediction_2026,ludwig_crystallization_2007,schleede_phase_2012} did not report a honeycomb crystal in mass-asymmetric \eh{} bilayers. The earlier path-integral Monte Carlo (PIMC) studies~\cite{ludwig_crystallization_2007,schleede_phase_2012} focused on $d\geq1$, whereas most of the honeycomb region identified here lies at smaller separations. At $d=1$, the honeycomb crystal is favored by only $\sim1.8$~mHa per electron, making it easily masked by method limitations. The PIMC calculations used Boltzmann statistics, omitting fermionic exchange, which can be important in the itinerant, strongly degenerate small-$\rs$ regime. Ref.~\cite{nashabeh_prediction_2026} used Hartree--Fock theory, which includes exchange but not correlation, and did not consider the honeycomb crystal as a candidate. Our calculations explicitly include this candidate and use the correlated NN-VMC method to settle the resulting millihartree-scale competition, identifying the honeycomb ground-state region.

Experimental realization of the honeycomb reconstruction requires a hierarchy of fast electrons and slow holes in an \eh{} bilayer, which a moiré potential can provide by trapping the holes while leaving the adjacent electron layer unpatterned~\cite{qi_exciton_2026}.
At an experimentally feasible \eh{} density $n_{\mathrm{phys}}=10^{11}~\text{cm}^{-2}$, an illustrative architecture comprising bilayer graphene (BLG), hexagonal boron nitride (hBN), and a WS$_2$/WSe$_2$ bilayer gives $\rs\approx2.78$, which also maps $d=0.05$--$1.0$ to $d_{\mathrm{phys}}\approx0.32$--$6.4~\text{nm}$ (End Matter Appendix~\ref{app:physical_unit_mapping}).
This range covers the $1.5$--$3.0$~nm hBN spacer thicknesses commonly used to balance the need for robust interlayer Coulomb coupling with the suppression of interlayer tunneling~\cite{ma_strongly_2021,qi_thermodynamic_2023,nguyen_perfect_2025,qi_perfect_2025,liu_high-temperature_2026,zhang_engineering_2024}.
This architecture also supplies the mass hierarchy assumed by the BO treatment. The BLG electron mass is $m_e^*\approx0.033\,m_e$~\cite{li_effective_2016}, whereas the WSe$_2$ valence-band mass is already an order of magnitude larger at $m_h^*\approx0.4\,m_e$~\cite{kormanyos_kp_2015}, yielding a mass ratio of $m_h^*/m_e^*>10$. The moiré potential further flattens the hole band, and correlated insulating states at commensurate filling pin the holes~\cite{regan_mott_2020,xu_correlated_2020}, effectively driving this mass ratio to an even larger value.

The predicted reconstruction can be tested through complementary measurements.
In real space, the triangular and honeycomb charge order can be directly distinguished using non-invasive STM with a graphene sensor~\cite{li_imaging_2021}, and
recent measurements have observed honeycomb and oblique electron crystals in rhombohedral hexalayer graphene~\cite{sun2026realspacevisualizationemergentelectron}.
As an independent reciprocal-space probe, the exciton Umklapp resonance could distinguish the two lattices~\cite{shimazaki_optical_2021,smolenski_signatures_2021,wang_spectroscopy_2026,qi_exciton_2026}, because the two lattices have different reciprocal vectors, $G_{\mathrm{HC}}=G_{\mathrm{T}}/\sqrt{2}$.
Thermodynamically, the crossing of the two energy-density branches produces a discontinuity in the chemical potential $\mu=\partial\mathcal E/\partial n$ and an anomaly in the inverse compressibility $\kappa^{-1}=n^2\partial\mu/\partial n$~\cite{zhang_signatures_2014,qi_thermodynamic_2023}, which can be accessed via asymmetric interlayer Coulomb drag~\cite{liu_high-temperature_2026}.
Finally, because each crystalline charge order coexists with several spin states, the spin sector requires separate probes, such as polarization-resolved optical spectroscopy~\cite{ciorciaro_kinetic_2023} and spin-dependent exciton Umklapp resonances~\cite{gomez_salvador_optical_2022}.

Beyond these static approximations, relaxing the constraint that fixes holes as static point charges opens pathways to more exotic phases, such as exciton condensates~\cite{zhu_two-dimensional_2025,luo_pairing-based_2026}, exciton supersolids~\cite{joglekar_wigner_2006}, and superconducting instabilities~\cite{nashabeh_prediction_2026}.
Finally, interstitial doping away from commensurability could promote a metallic Wigner crystal~\cite{kim_dynamical_2024,han_evidence_2026}, while the inclusion of Berry curvature effects may further broaden this intermediate regime and induce anomalous Hall crystal phases~\cite{miao_emergence_2026}.

\begin{acknowledgments}
    \paragraph{Acknowledgments} We thank colleagues at Interdisciplinary Center for Theoretical Physics and Information Sciences and the Shanghai Artificial Intelligence Laboratory for helpful discussions.
    This work is supported by the National Natural Science Foundation of China under Grants No. T2225018, No. 12188101, No. T2121001, No. 12634001, the Cross-Disciplinary Key Project of Beijing Natural Science Foundation No. Z250005, New Generation Artificial Intelligence-National Science and Technology Major Project (Grants No. 2025ZD0121802), Quantum Science and Technology-National Science and Technology Major Project (Grants No. 2021ZD0302400).
    X.D. is supported by a fellowship award from Hong Kong Research Grant Council (Project No. SRFS2324-6S01) and New Conerstone Science Foundation.
    Calculations were performed using approximately 7,000 NVIDIA H200 GPU-hours in total.
\end{acknowledgments}

\bibliography{paper}
\onecolumngrid

\begin{center}
    \textbf{End Matter}
\end{center}
\vspace{2mm plus 2mm minus 2mm}
\twocolumngrid

\setcounter{secnumdepth}{3}
\appendix

\section{Charge and spin structure factors}
\label{app:localization_diagnostics}

We characterize charge localization and spin order through the Fourier modes
\begin{equation}
    \rho_c(\mathbf q)=\sum_j e^{-i\mathbf q\cdot\mathbf e_j},
    \qquad
    \rho_s(\mathbf q)=\sum_j\xi_j e^{-i\mathbf q\cdot\mathbf e_j},
    \label{eq:charge_spin_modes}
\end{equation}
where $\xi_j=+1$ for spin-up and $-1$ for spin-down electrons.
The equal-time charge and spin structure factors are
\begin{equation}
    S_\alpha(\mathbf q)=\frac{1}{N}\la|\rho_\alpha(\mathbf q)|^2\ra,
    \qquad \alpha\in\{c,s\}.
    \label{eq:charge_spin_structure}
\end{equation}
The coherent spin structure factor is
\begin{equation}
    S_{s,\rm coh}(\mathbf q)
    =\frac{|\langle\rho_s(\mathbf q)\rangle|^2}{N}.
    \label{eq:coherent_spin_structure}
\end{equation}
We reduce each two-dimensional structure factor to a radial spectrum by taking the maximum over the nonzero shell $\mathcal Q_q=\{\mathbf k:|\mathbf k|=q\}$:
\begin{equation}
    \begin{aligned}
        S_\beta^{\max}(q)
         & =\max_{\mathbf k\in\mathcal Q_q}S_\beta(\mathbf k),                                      \\
        \widetilde S_\beta^{\max}(q)
         & =\frac{S_\beta^{\max}(q)}{\max_{q'>0}S_\beta^{\max}(q')}, \beta \in \{s,c,(s,\rm coh)\}.
    \end{aligned}
    \label{eq:radial_shell_spectra}
\end{equation}
In Fig.~\ref{fig:magnetic_textures}, they are plotted against $q/q_{\min}$, where $q_{\min}=\min_{\mathbf q\ne0}|\mathbf q|$.

The stripe ordering wave vector is
$
    \mathbf Q_\star
    =\operatorname*{arg\,max}_{\mathbf q\neq0}S_{s,\rm coh}(\mathbf q).
$
For every stripe state, this maximum lies along $x$ at $Q_x/q_{\min,x}=2$, where $q_{\min,x}$ is the magnitude of the smallest nonzero reciprocal vector along $x$.
The corresponding stripe width is therefore
$
    w=\frac{\pi}{Q_x}=\frac{L_x}{4},
$
one quarter of the simulation-cell length along $x$.

\section{Mapping to experimental scales}
\label{app:physical_unit_mapping}

Within the isotropic constant-$\epsilon_r$ model, the effective Bohr radius $a_{\mathrm{B}}^*=\epsilon_r(m_e/m_e^*)a_{\mathrm{B}}$ sets the physical length scale, giving $n_{\mathrm{phys}}=[\pi(\rs a_{\mathrm{B}}^*)^2]^{-1}$ and $d_{\mathrm{phys}}=d a_{\mathrm{B}}^*$.
An illustrative BLG parameter set with $m_e^*=0.033\,m_e$~\cite{li_effective_2016} and $\epsilon_r=4$ gives $a_{\mathrm{B}}^*\approx6.41~\text{nm}$.
At $n_{\mathrm{phys}}=10^{11}~\text{cm}^{-2}$, this parameter set gives $\rs\approx2.78$.
The interval $d=0.05$--$1.0$, in which the honeycomb crystal emerges, then corresponds to $d_{\mathrm{phys}}\approx0.32$--$6.4~\text{nm}$.

\section{Hubbard parameters for the honeycomb hole lattice}
\label{app:electronic_character}

We estimate the nearest-neighbor hopping $t$ and on-site Coulomb
repulsion $U$ using a set of orthonormal localized orbitals
centered at the fixed hole sites.

\paragraph{Localized orbitals.}
We characterize electron localization on the honeycomb hole lattice by
\begin{equation}
    \begin{aligned}
        f_h(\mathbf r)
         & = \min_I |\mathbf r-\mathbf h_I|^2, \\
        \xi_{\rm VMC}^2
         & = \frac{1}{N}
        \left\langle \sum_{a=1}^N f_h(\mathbf e_a) \right\rangle,
    \end{aligned}
    \label{eq:hubbard_nearest_distance}
\end{equation}
where the minimum includes all periodic images of the holes.

At each hole site
$\mathbf R_i=\mathbf R+\boldsymbol\tau_\alpha$,
where $\mathbf R$ is a Bravais lattice vector and
$\alpha=A,B$ labels the two sublattices, we introduce a normalized
Gaussian function with trial width $\xi$:
\begin{equation}
    \begin{aligned}
        g_i(\mathbf r;\xi)
         & = \frac{1}{\sqrt{\pi}\,\xi}
        e^{-|\mathbf r-\mathbf R_i|^2/(2\xi^2)}, \\
        S_{ij}(\xi)
         & = \braket{g_i|g_j}
        = e^{-|\mathbf R_i-\mathbf R_j|^2/(4\xi^2)}.
    \end{aligned}
    \label{eq:hubbard_gaussian_overlap}
\end{equation}
Symmetric L\"owdin
orthogonalization~\cite{lowdin_nonorthogonality_1950}
then gives
\begin{equation}
    w_i(\mathbf r;\xi)
    = \sum_j g_j(\mathbf r;\xi)(S^{-1/2})_{ji},
    \label{eq:hubbard_lowdin}
\end{equation}
with $\braket{w_i|w_j}=\delta_{ij}$.

Assuming one electron per site and neglecting off-diagonal elements
of the one-body density matrix, we approximate the spin-summed
electron density by
\begin{equation}
    n(\mathbf r;\xi)
    = \sum_{\mathbf R,\alpha=A,B}
    |w_{\mathbf R\alpha}(\mathbf r;\xi)|^2.
    \label{eq:hubbard_density_model}
\end{equation}
Each orbital is normalized over the full plane, so the resulting
periodic density satisfies
$\int_{\rm cell}n(\mathbf r;\xi)\,d^2r=2$
for the two-site primitive cell.

We choose $\xi_*$ such that this density reproduces the NN-VMC
localization measure:
\begin{equation}
    F(\xi_*)=\xi_{\rm VMC}^2,
    \qquad
    F(\xi)
    = \frac{1}{2}\int_{\rm cell}d^2r\,
    n(\mathbf r;\xi)f_h(\mathbf r).
    \label{eq:hubbard_width_matching}
\end{equation}

\paragraph{Nearest-neighbor hopping.}
The single-electron Hamiltonian is
$h_0=-\nabla^2/2+\hat{v}_{eh}(\mathbf r)$, where $\hat{v}_{eh}$ is the interlayer electron--hole potential of the fixed hole lattice, so that summing $h_0$ over electrons gives $\hat T+\hat V_{eh}$.
The Fourier coefficients of $\hat{v}_{eh}$ are
\begin{equation}
    V_{\mathbf G}
    = -\frac{2\pi}{A_c}\frac{e^{-Gd}}{G}
    \sum_{\alpha=A,B}
    e^{-i\mathbf G\cdot\boldsymbol\tau_\alpha},
    \qquad G=|\mathbf G|>0,
    \label{eq:hubbard_hole_potential}
\end{equation}
where $A_c=2\pi\rs^2$ is the primitive-cell area.

With $R_{ij}=|\mathbf R_i-\mathbf R_j|$ and
$\mathbf M_{ij}=(\mathbf R_i+\mathbf R_j)/2$,
the matrix elements of $h_0$ in the Gaussian basis are
\begin{equation}
    \begin{aligned}
        h^{(g)}_{ij}
         & = \braket{g_i|h_0|g_j}                       \\
         & = S_{ij}\left[
                       \frac{1}{2\xi^2}
                       - \frac{R_{ij}^2}{8\xi^4}
                       + \overline V_\xi(\mathbf M_{ij})
                       \right],
    \end{aligned}
    \label{eq:hubbard_gaussian_hamiltonian}
\end{equation}
where
\begin{equation}
    \overline V_\xi(\mathbf x)
    = \sum_{\mathbf G\ne0}V_{\mathbf G}
    e^{-\xi^2G^2/4}e^{i\mathbf G\cdot\mathbf x}
    \label{eq:hubbard_smoothed_potential}
\end{equation}
is $\hat{v}_{eh}$ averaged over a Gaussian density of width
$\xi$ centered at $\mathbf x$.

The Hamiltonian in the orthonormal basis and the nearest-neighbor
hopping are
\begin{equation}
    \begin{aligned}
        h^{(w)}
         & = S^{-1/2}h^{(g)}S^{-1/2},           \\
        t
         & = -h^{(w)}_{0A,0B}\big|_{\xi=\xi_*},
    \end{aligned}
    \label{eq:hubbard_orthogonal_hopping}
\end{equation}
where $0A$ and $0B$ denote a nearest-neighbor pair.

\paragraph{On-site Coulomb repulsion.}
Using the same orthonormal orbitals at the calibrated width
$\xi_*$, we evaluate
\begin{equation}
    U
    = \iint d^2r\,d^2r'\,
    \frac{
        |w_{0A}(\mathbf r;\xi_*)|^2
        |w_{0A}(\mathbf r';\xi_*)|^2
    }{|\mathbf r-\mathbf r'|}.
    \label{eq:hubbard_onsite_integral}
\end{equation}
With $C=S(\xi_*)^{-1/2}$, direct integration using
Eq.~(\ref{eq:hubbard_lowdin}) gives
\begin{equation}
    \begin{aligned}
        U
         & = \sum_{ijkl} C_{i,0A}C_{j,0A}C_{k,0A}C_{l,0A} \\
         & \quad\times S_{ij}(\xi_*)S_{kl}(\xi_*)
        J_{\xi_*}(|\mathbf M_{ij}-\mathbf M_{kl}|),       \\
        J_\xi(L)
         & = \frac{\sqrt{\pi/2}}{\xi}
        e^{-x}I_0(x),
        \qquad x=\frac{L^2}{4\xi^2},
    \end{aligned}
    \label{eq:hubbard_gaussian_coulomb_sum}
\end{equation}
where $I_0$ is the modified Bessel function of the first kind.

\end{document}